\documentclass[10pt,journal]{IEEEtran}
\usepackage{subfigure}
\usepackage{graphicx}
\usepackage{latexsym}
\usepackage{psfrag}
\usepackage{color}
\usepackage[cmex10]{amsmath}
\usepackage{amstext,amssymb, amsfonts}
\usepackage{chngcntr}
\def\sinc{\mbox{sinc}}

\def\rec{\mathsf{r}}
\def\sam{\mathsf{s}}
\def\rect{\mbox{box}_{\Delta_b}}
\DeclareMathOperator{\BigO}{O}

\def\mindex#1{\index{#1}}

\def\sq{\hbox{\rlap{$\sqcap$}$\sqcup$}}
\def\qed{\ifmmode\sq\else{\unskip\nobreak\hfil
\penalty50\hskip1em\null\nobreak\hfil\sq
\parfillskip=0pt\finalhyphendemerits=0\endgraf}\fi\medskip}

\long\def\defbox#1{\framebox[.9\hsize][c]{\parbox{.85\hsize}{%
\parindent=0pt
\baselineskip=12pt plus .1pt      
\parskip=6pt plus 1.5pt minus 1pt 
 #1}}}

\long\def\beginbox#1\endbox{\subsection*{}%
\hbox{\hspace{.05\hsize}\defbox{\medskip#1\bigskip}}%
\subsection*{}}

\def\endbox{}

\newsavebox{\junk}
\savebox{\junk}[1.6mm]{\hbox{$|\!|\!|$}}

\newcommand{\field}[1]{\mathbb{#1}}

\def\Re{\field{R}}

\def\bZ{{\mathbb Z}}

\def\bfmath#1{{\mathchoice{\mbox{\boldmath$#1$}}%
{\mbox{\boldmath$#1$}}%
{\mbox{\boldmath$\scriptstyle#1$}}%
{\mbox{\boldmath$\scriptscriptstyle#1$}}}}

\def\bfmY{\bfmath{Y}}

\def\bfmhhaY{\bfmath{\hhaY}} 
\def\bfmhhaY{\hbox to 0pt{$\widehat{\bfmY}$\hss}\widehat{\phantom{\raise 1.25pt\hbox{$\bfmY$}}}}

\def\til={{\widetilde =}}

\def\clI{{\cal I}}

 \def\FRAC#1#2#3{\genfrac{}{}{}{#1}{#2}{#3}}

\def\ddtp{{\mathchoice{\FRAC{1}{d^{\hbox to 2pt{\rm\tiny +\hss}}}{dt}}%
{\FRAC{1}{d^{\hbox to 2pt{\rm\tiny +\hss}}}{dt}}%
{\FRAC{3}{d^{\hbox to 2pt{\rm\tiny +\hss}}}{dt}}%
{\FRAC{3}{d^{\hbox to 2pt{\rm\tiny +\hss}}}{dt}}}}

\def\half{{\mathchoice{\FRAC{1}{1}{2}}%
{\FRAC{1}{1}{2}}%
{\FRAC{3}{1}{2}}%
{\FRAC{3}{1}{2}}}}

\def\average#1,#2,{{1\over #2} \sum_{#1}^{#2}}

\def\eye(#1){{\bf(#1)}\quad}

\newtheorem{theorem}{Theorem}[section]

\newtheorem{proposition}[theorem]{Proposition}

\def\eq#1/{(\ref{e:#1})}

\newcommand{\beqn}[1]{\notes{#1}%
\begin{eqnarray} \elabel{#1}}

\newcommand{\eeqn}{\end{eqnarray} }

\newcommand{\beq}[1]{\notes{#1}%
\begin{equation}\elabel{#1}}

\newcommand{\eeq}{\end{equation}}

\def\bdes{\begin{description}}
\def\edes{\end{description}}

\newcounter{rmnum}
\newenvironment{romannum}{\begin{list}{{\upshape (\roman{rmnum})}}{\usecounter{rmnum}
\setlength{\leftmargin}{14pt}
\setlength{\rightmargin}{8pt}
\setlength{\itemindent}{-1pt}
}}{\end{list}}

\newcounter{anum}

{\end{list}}

\def\ass(#1:#2){(#1\ref{#1:#2})}

\def\ritem#1{
\item[{\sf \ass(\current_model:#1)}]
}

\newenvironment{recall-ass}[1]{%
\begin{description}
\def\current_model{#1}}{
\end{description}
}

\newcommand{\bd}{\begin{description}}
\newcommand{\ed}{\end{description}}
\newcommand{\bt}{\begin{theorem}}
\newcommand{\et}{\end{theorem}}
\newcommand{\ba}{\begin{array}{rcl}}
\newcommand{\ea}{\end{array}}

\newlength{\noteWidth}
\long\def\notes#1{\ifinner
             {\tiny #1}
             \else
              \marginpar{\parbox[t]{\noteWidth}{\raggedright\tiny #1}}
               \fi}

\counterwithin*{corollary}{theorem}
\usepackage{amssymb}
\usepackage{pifont}
\newcommand{\cmark}{\ding{51}}%
\newcommand{\xmark}{\ding{55}}%
\begin{document}
\title{Sampling and Reconstruction of Spatial Fields using Mobile Sensors}
\author{\authorblockN{Jayakrishnan Unnikrishnan and Martin Vetterli\\}
\authorblockA{Audiovisual Communications Laboratory,
School of Computer and Communication Sciences,\\
Ecole Polytechnique F\'{e}d\'{e}rale de Lausanne (EPFL),
Switzerland\\
Email:  \{jay.unnikrishnan, martin.vetterli\}@epfl.ch}
}
\maketitle
\begin{abstract}
Spatial sampling is traditionally studied in a static setting where static sensors scattered around space take measurements of the spatial field at their locations. In this paper we study the emerging paradigm of sampling and reconstructing spatial fields using sensors that move through space. We show that mobile sensing offers some unique advantages over static sensing in sensing time-invariant bandlimited spatial fields. Since a moving sensor encounters such a spatial field along its path as a time-domain signal, a time-domain anti-aliasing filter can be employed prior to sampling the signal received at the sensor. Such a filtering procedure, when used by a configuration of sensors moving at constant speeds along equispaced parallel lines, leads to a complete suppression of spatial aliasing in the direction of motion of the sensors. We analytically quantify the advantage of using such a sampling scheme over a static sampling scheme by computing the reduction in sampling noise due to the filter. We also analyze the effects of non-uniform sensor speeds on the reconstruction accuracy. Using simulation examples we demonstrate the advantages of mobile sampling over static sampling in practical problems.

We extend our analysis to sampling and reconstruction schemes for monitoring time-varying bandlimited fields using mobile sensors. We demonstrate that in some situations we require a lower density of sensors when using a mobile sensing scheme instead of the conventional static sensing scheme. The exact advantage is quantified for a problem of sampling and reconstructing an audio field.
\end{abstract}

\section{Introduction}

The typical approach for measuring a spatial field makes use of static sensors distributed over the area of interest \cite{ingbarschvetcoupar10}. Consequently much of the literature on spatial sampling and reconstruction have focused on such static sensing schemes \cite{kumishram11}, \cite{reimat09}. An emerging paradigm in spatial sampling is the use of mobile sensors that move through the area of interest, taking measurements along their paths \cite{sinnowram06}, \cite{ciglurripvan08}. Mobile sensing schemes have several advantages over static schemes, the chief of which is the fact that a single mobile sensor can be used to take measurements at several distinct positions in space. Such a scheme is often more cost-effective and easier to implement than static sensing since it requires only a single sensor for monitoring a large spatial area. In this paper we illustrate and analyze various unique aspects of mobile sensing and the advantages that they offer.

Consider a time-invariant spatial field\footnote{A field that is a function of space alone and does not vary with time.} defined over $d$-dimensional space represented by a square integrable mapping $f: \Re^d \mapsto \Re$ with $f \in L^2(\Re^d)$. For any $r \in \Re^d$, the quantity $f(r)$ represents the value of the field at spatial location $r$. The field could represent, for instance, some spatially varying parameter like the temperature of air or the concentration of a pollutant in the air. Suppose further that the field $f$ is slowly varying in space and can hence be modeled as a spatially bandlimited field. Let $\tilde f$ represent the observed field that is a noisy version of the field of interest expressed as
\begin{equation}
\tilde f(r) = f(r) + w(r), r \in \Re^d \label{eqn:field}
\end{equation}
where $w$ denotes non-bandlimited spatial noise, which we refer to as environmental noise. The objective in a typical sampling and reconstructing scheme is to use the samples of the observed field $\tilde f$ to obtain a reconstruction $\hat f$ of the field $f$ such that the mean-square error (MSE) $\mathsf{E}[\|\hat f - f\|_2^2]$ in the reconstruction is minimal. If the noise $w$ is absent, then the observed field $\tilde f =f$ is bandlimited in space, and we know from classical sampling theory \cite{petmid62} that we can recover the field exactly from samples of the field taken by static sensors located on a lattice of points in space. However, if $w \neq 0$ then the observed field $\tilde f$ is not bandlimited and we expect to see some effects of spatial aliasing while sampling the field using static sensors. More importantly, unlike in the case of sampling a time-domain signal, there is no way to implement a spatial anti-aliasing filter in a static sensing setup. This drawback of spatial sensing using static sensors has been observed in various applications (see, e.g., 
\cite{rafweibac07}, \cite{kumishram10}, \cite{sporab06}).

Sampling using mobile sensors provides us with an approach that partially addresses the issue of spatial aliasing. 
Mobile sensing allows filtering in time prior to sampling which induces filtering over space in the direction of motion of the sensor. Such spatial filtering is not possible in a static sensing setup. A moving sensor receives as input a time-domain signal representing the field along the path of the sensor 
given by,
\begin{equation}
\tilde s(t) =\tilde f(r(t))= f(r(t)) + w(r(t))\label{eqn:timewrapdsignal}
\end{equation}
where $r(t) \in \Re^d$ denotes the position of the sensor at time $t$. The analog signal $\tilde s(t)$ can be passed through an analog anti-aliasing filter \emph{prior} to discretizing into samples. Such filtering discards out-of-band noise in the direction of motion of the sensor thus inducing \textit{spatial smoothing}. Implementing such a filter requires redesigning of the sensing process employed in the sensor and may be feasible only in some scenarios. For the purpose of illustration consider a problem of measuring the concentration of a gas in the air along a straight road, assumed to be constant in time. Assume further that one has a sensor that can measure the average concentration of the gas in a chamber. We want to design a scheme that computes the filtered samples
\[
s_n := \int_{-\infty}^\infty \tilde s(\tau) h(nT - \tau) d\tau
\]
at times $nT$ for all $n \in \bZ$. If we assume that $h$ is a finite impulse response filter satisfying $h(t) \geq 0$ for all $t$ and $h(t) = 0$ for $|t| \geq \frac{T}{2}$, then such a filter can be implemented by designing a system in which a chamber is mounted on a vehicle moving along the road at a constant velocity. Air is pumped into the chamber through an opening whose size can be varied dynamically such that the rate of entry of air at time $\tau$ is proportional to $h(T \lfloor \frac{\tau }{T} + \half \rfloor  - \tau)$ where $\lfloor x \rfloor$ indicates integer component of a real number $x$. The chamber is periodically emptied after each sample is measured at times $(n+\half)T$ for all $n \in \bZ$. Such a scheme essentially implements an analog domain sampling prefilter with impulse response $h$ up to a normalization constant. This scheme can be generalized to more general non-negative finite impulse responses $h$ with wider support, if we use multiple chambers.

\begin{figure*}
\centering
\subfigure[Spectrum of field bandlimited to $\Re\times{[-\rho,\rho]}$.]{
\psfrag{oy}{$\omega_y$}
\psfrag{ox}{$\omega_x$}
\psfrag{Hom0}{$H_{\sam}(\omega)=0$}
\psfrag{Hom1}{$H_{\sam}(\omega)=1$}
\psfrag{r}{$\rho$}
\psfrag{mr}{$-\rho$}
\psfrag{Fz}{$\scriptstyle{F(\omega) = 0}$}
\includegraphics[width=2.2in]{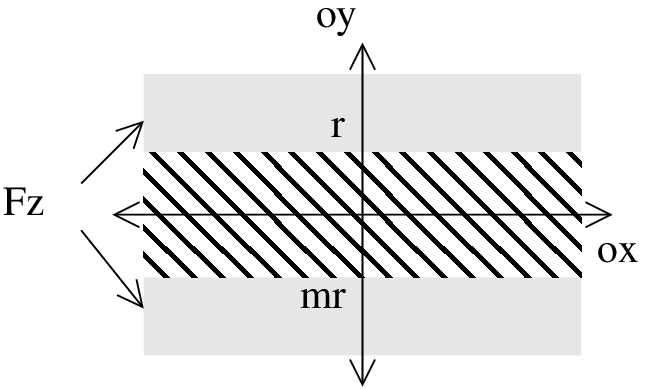}
\label{fig:specBLwy}
}
\subfigure[Aliased spectrum under static sampling.]{
\psfrag{oy}{$\omega_y$}
\psfrag{ox}{$\omega_x$}
\psfrag{Hom0}{$H_{\sam}(\omega)=0$}
\psfrag{Hom1}{$H_{\sam}(\omega)=1$}
\psfrag{r}{$\rho$}
\psfrag{mr}{$-\rho$}
\psfrag{Wy}{$\frac{2\pi}{\Delta_y}$}
\psfrag{A}[rt][rt]{\shortstack[l]{{\small Reconstruction}\\{\small is aliased}}}
\includegraphics[width=2.1in]{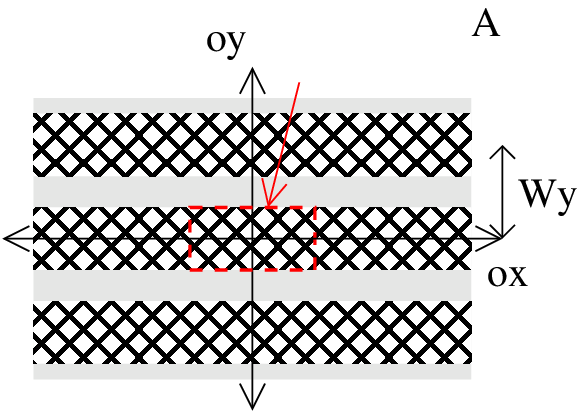}
\label{fig:specBLwystat}
}
\subfigure[No aliasing under mobile sampling with filtering in the $x$-direction.]{
\psfrag{oy}{$\omega_y$}
\psfrag{ox}{$\omega_x$}
\psfrag{Hom0}{$H_{\sam}(\omega)=0$}
\psfrag{Hom1}{$H_{\sam}(\omega)=1$}
\psfrag{r}{$\rho$}
\psfrag{mr}{$-\rho$}
\psfrag{Wy}{$\frac{2\pi}{\Delta_y}$}
\psfrag{Wx}{$\frac{2\pi}{\Delta_x}$}
\psfrag{U}[rt][rt]{\shortstack[l]{{\small Unaliased}\\
{\small reconstruction}}}
\includegraphics[width=2.1in]{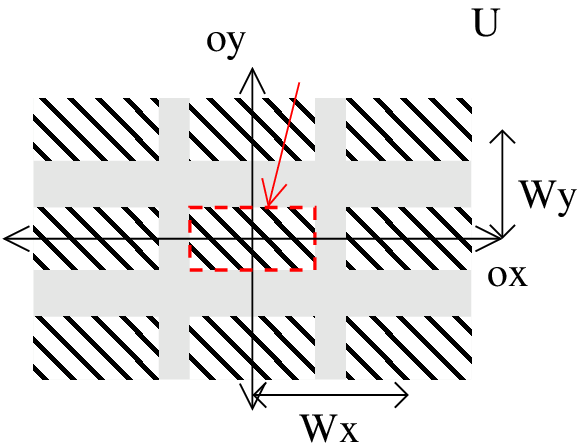}
\label{fig:specBLwymbl}
}
\caption[Optional caption for list of figures]{Sampling a two-dimensional field bandlimited only in one direction. Spectrum under static sampling is aliased but that under mobile sampling is not.}
\label{fig:specBLwyall}
\end{figure*}

However, in the proposed approach, a caveat to note is the peculiar fact that such filtering permits spatial smoothing only in the direction of motion of the sensor. Hence, for spatial fields of dimension $d > 1$, the anti-aliasing filters are \textit{thin}, i.e., the effective spatial impulse response of the filter is supported on a set of dimension $1$. Thus this form of spatial smoothing allows us to discard only one component of the out-of-band noise. For the sake of illustrating the potential advantage offered by such a filtering scheme, let us consider the problem of sampling a field in two-dimensional space having a Fourier transform that is bandlimited only in one direction as shown in Figure \ref{fig:specBLwy}. Sampling such a field using static sensors will always lead to aliasing because the repetitions in the sampled spectra necessarily overlap. An example of the sampled spectrum obtained by static sampling on a lattice of the form $\{(m\Delta_x,n\Delta_y):m,n\in\bZ\}$ is shown in Figure \ref{fig:specBLwystat} where $\Delta_y < \frac{2\pi}{\rho}$. However, we will see later that such a field can be sampled on the same lattice using sensors that move along equispaced straight lines parallel to the $x$-axis as shown in Figure \ref{fig:unifset} and use ideal anti-aliasing filters in the time domain. This leads to a complete suppression of aliasing as shown in Figure \ref{fig:specBLwymbl}. We analyze such a mobile sensing scheme for sampling bandlimited fields later in the paper and quantify advantages obtained in terms of suppressing out-of-band noise.

The scenario is a little different in the case of sampling time-varying bandlimited fields, represented by functions of both time and space of the form $f(r,t)$ where $r$ denotes position and $t$ time. Here the advantages of mobile sensing are less pronounced since the field values at various points in space are also varying in time. Hence it is possible to filter in time even with static sensors. Furthermore, one has to account for the Doppler effect while sensing with moving sensors (see e.g., \cite[sec 5.2]{ajd06}). Nevertheless we show that in some scenarios mobile sensing requires fewer sensors than the number of sensors required with static sampling. These results could have potential applications in limiting spatial aliasing in certain sampling problems, e.g., wave-field synthesis \cite{sporab06}.


The advantage of mobile sensing over static sensing in suppressing spatial aliasing has been noted in the context of audio source localization 
in \cite{ciglurripvan08} where the authors show that using a planar rotating array of microphones, the effective number of measurement points can be increased thus reducing the amount of spatial aliasing while  using the beamforming technique. In some other works moving microphones have been used for estimating room impulse responses \cite{ajdsbavet07} and  head related impulse responses \cite{enz08}. In these works the desired responses are estimated from a finely sampled version of the audio signal received at the moving microphones. Other works on mobile sensing focus on adaptive path-planning algorithms \cite{sinnowram06}, \cite{sriram06} for environmental monitoring. However, to the best of our knowledge, this is the first work to illustrate the possibility of implementing spatial smoothing by using a mobile sensing scheme together with time-domain anti-aliasing filtering, and to quantify the improvements of such a scheme over static sensing. In earlier work \cite{unnvet11TITb}, we studied the problem of designing trajectories for mobile sensing which minimize the total distance required to be traveled by the sensors per unit area of the field being sampled.

The paper is organized as follows. We study the sampling of time-invariant fields in Section \ref{sec:timeinv} and time-varying fields in Section \ref{sec:tvarying}. We discuss sensor trajectories, filter designs, reconstruction schemes, and comparisons with static sensing in various aspects. In order to highlight various advantages of mobile sensing we have considered several different models for the field and the noise. To enable easy navigation through the paper we summarize the model assumptions used in the various subsections in Table \ref{tbl:secs}. In Section \ref{sec:sims} we discuss a simulation example comparing mobile and static sampling for a practical problem and conclude in Section \ref{sec:conc}.

\begin{table}
    \begin{tabular}{ | p{0.33in} | p{0.58in} | p{0.61in} | p{0.55in} |p{0.55in} |}
    \hline
Section&\multicolumn{2}{|c|}{Field properties} & Noise added& Noise added\\ \cline{2-3}
 & Bandlimited & Time-varying &to the field&to samples\\ \hline
II.A &$\qquad$ \cmark &$\qquad$ \xmark &$\qquad$ \xmark &$\qquad$ \xmark \\ \hline
II.B,D,F &$\qquad$ \cmark &$\qquad$ \xmark &$\qquad$ \cmark &$\qquad$ \xmark \\ \hline
II.C &$\qquad$ \xmark &$\qquad$ \xmark &$\qquad$ \xmark &$\qquad$ \xmark \\ \hline
II.E &$\qquad$ \cmark &$\qquad$ \xmark &$\qquad$ \xmark &$\qquad$ \cmark \\ \hline
III.A,B &$\qquad$ \cmark &$\qquad$ \cmark &$\qquad$ \xmark &$\qquad$ \xmark \\ \hline
    \end{tabular}
        \caption{Summary of model assumptions in different sections. The last two columns represent two different additive noise models that we consider - analog spatial noise added to the field prior to sampling, and discrete noise added to the field measurements obtained after sampling.}\label{tbl:secs}
\end{table}


\section{Time-invariant fields}\label{sec:timeinv}
Consider a time-invariant field in $d$-dimensional space represented by a mapping $f: \Re^d \mapsto \Re$ with $f \in L^2(\Re^d)$. We define its Fourier transform $F$ as
\[
F(\omega) = \int_{\Re^d} f(r) \exp(-{\sf i}\langle \omega, r\rangle ) dr, \qquad \omega \in \Re^d
\]
where ${\sf i}$ denotes the imaginary unit, and $\langle u ,v \rangle$ denotes the Euclidean inner product between vectors $u$ and $v$. We assume that the field $f$ is bandlimited to a set $\Omega \subset \Re^d$, i.e., suppose that the Fourier transform $F$ of $f$ is supported on a known set $\Omega \subset \Re^d$, so that $F(\omega) = 0$ for $\omega \notin \Omega$. We use $\tilde f$ to denote the observed field, which may either be the field $f$ itself or a noisy version of it as shown in (\ref{eqn:field}). In the noisy case, we model the environmental noise $w(r)$ as a zero mean wide sense stationary (WSS) process with unknown power spectral density $\mathsf{S}_w$. For mathematical regularity, we assume that the noise power spectral density decays at a faster rate than $O(\|\omega\|_2^{-d})$, i.e., we assume that
\begin{equation}
\mathsf{S}_w(\omega) = O(\|\omega\|_2^{-r}) \mbox{ for some }r > d. \label{eqn:noisemodel}
\end{equation}
Time-invariant fields are particularly well-suited for mobile sensing schemes since the field does not vary as the sensor moves around taking measurements in space. We distinguish between two distinct scenarios - one-dimensional fields in which a single moving sensor can visit \emph{all} points in the one-dimensional spatial region of interest, and higher-dimensional fields in which each sensor can measure the field only on a one-dimensional sub-manifold of the spatial region of interest. To illustrate this difference we consider one-dimensional fields and two-dimensional fields with the understanding that the analysis for two-dimensional fields can be easily generalized to higher-dimensional fields. In this section, we detail the mobile sensing scheme and describe how a time domain anti-aliasing filter can be used to perform spatial smoothing. We quantify the improvements of such a sensing scheme over the static sensing scheme. We initially assume that the sensors are moving with constant velocities and later consider the scenario with non-uniform speeds.

\subsection{Sensor trajectories for mobile sensing}
For sampling a field in $\Re^d$ where $d \geq 2$, there are several possible choices of trajectories that can be used by the moving sensors. In our recent work \cite{unnvet11TITb} we studied the problem of designing sensor trajectories that admit perfect reconstruction of bandlimited fields from measurements taken by the sensors moving along these trajectories in a noise-free setting. We introduced the notion of optimal trajectories that minimize the total distance required to be traveled by the moving sensors. For $d=2$ we showed that a set of trajectories comprising one set of equispaced parallel lines is optimal from certain classes of trajectories. In this paper, we will restrict ourselves to such a collection of trajectories for fields in $\Re^2$ and assume that we have one sensor moving along each line taking measurements on its path.

\subsection{Sampling and reconstruction}
Consider a sensor moving at a constant velocity through space. The position $r(t)$ of the sensor at time $t$ is given by an affine function of the form $r(t) = u + vt$ where $u,v \in \Re^d$ represent the initial position and velocity of the sensor respectively. Without loss of generality we assume that $u = 0$ for simplifying the analysis. In the absence of noise the time-domain signal seen by such a sensor is given by
\[
s_0(t) = f(r(t)) = f(u+vt), t\in\Re.
\]
In this scenario it easily follows (see, e.g., \cite[Lemma 2.2]{unnvet11TITb}) that the signal $s_0(.)$ is bandlimited to
\begin{equation}
\Omega_{s_0} := \{\langle v , \omega\rangle: \omega \in \Omega\} \subset \Re.\label{eqn:s0BW}
\end{equation}
It follows via the Nyquist sampling theorem that $s_0(.)$ can be perfectly recovered by sampling it uniformly at temporal intervals less than or equal to ${2\pi}/({\max \Omega_{s_0} - \min \Omega_{s_0}})$. Furthermore, in the presence of noise, the signal received by the sensor can be passed through an anti-aliasing filter with passband aligned with $\Omega_{s_0}$ prior to sampling. This limits the contribution of out-of-band noise in the samples.

\subsubsection{One-dimensional field}\label{sec:samrecon1d}
Suppose $f(.)$ denotes a one-dimensional field bandlimited to $\Omega = [-\rho, \rho]$. In this case, if a sensor moves along the field at a constant velocity $v$, the signal it sees in the absence of noise is bandlimited to $[-v \rho, v \rho]$ as we argued in (\ref{eqn:s0BW}). Hence, in the presence of noise, the signal can be filtered prior to sampling. Let $h(.)$ denote the impulse response of an ideal filter with passband in the interval $[- \rho, \rho]$:
\begin{equation}
h(x) = \frac{\rho }{\pi}\sinc(\frac{\rho x}{\pi}), \quad x \in \Re\label{eqn:hfilter}
\end{equation}
where $\sinc(x) := \frac{\sin \pi x}{\pi x}$. Then the ideal choice for an anti-aliasing filter while sampling the time-domain signal $\tilde s(t) = \tilde f(vt)$ is given by $h_{\mathsf{aa}}(t) = v h(vt)$. Let $s(t)$ denote the signal at the output of the anti-aliasing filter. Suppose the sensor takes measurements every $T$ time units after passing the observed signal through the filter. Thus we get uniform samples of
\[
s(t) = (\tilde s* h_{\mathsf{aa}})(t) = (s_0 * h_{\mathsf{aa}})(t) + (\check w * h_{\mathsf{aa}})(t)
\]
where $\check w(t) = w(vt)$ and $*$ denotes convolution. Since $f$ is bandlimited we can write
\[
s(t) = f(v t) + (w * h)(v t).
\]
Now, if the sensor takes samples every $T$ time units, we essentially get uniform noisy samples of the field at intervals of $vT$ spatial units. We know from classical sampling theory \cite{oppschbuc99} that when noise is absent the field $f(.)$ can be exactly recovered from these samples provided that the sampling interval satisfies $T < \frac{\pi}{v \rho}$. Furthermore, in the noisy case, out-of-band noise can be suppressed in the reconstruction of the field by using sinc interpolation\footnote{Note that our rationale behind using sinc filters for anti-aliasing and interpolation is the fact that the only assumption we have about the field is that it is bandlimited. For a stochastic field model the MSE can be minimized by using a Wiener filter \cite{ramvanbluuns08}.},
\begin{equation}
\hat f(x) = \sum_{j \in \bZ} \frac{v T \rho }{\pi} s(j T) \sinc\left(\frac{\rho(x - j vT)}{\pi}\right),  \quad x \in \Re\label{eqn:recon1d}
\end{equation}
provided $T < \frac{\pi}{v \rho}$. This interpolation is well-defined when the noise satisfies (\ref{eqn:noisemodel}). However sinc interpolation is sensitive to errors in the samples and can lead to unbounded errors in the approximation at some values of $x$. This can be avoided by using alternate kernels in place of the sinc kernel (see, e.g., \cite{cvedaulog07} and references therein).
\begin{figure*}
\centering
\subfigure[Equispaced parallel line trajectories.]{
\psfrag{x}{$x$}
\psfrag{y}{$y$}
\psfrag{v}{$v$ (velocity)}
\psfrag{D}{$\Delta_y$}
\includegraphics[width=1.8in]{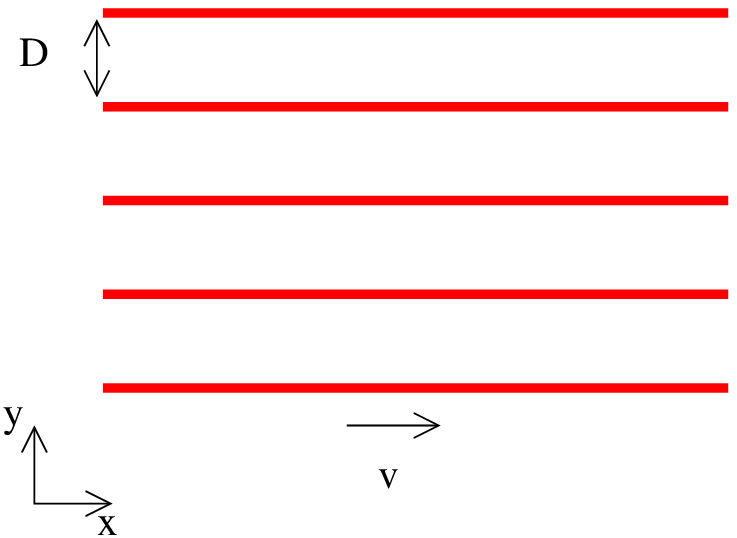}
\label{fig:unifset}
}
\subfigure[Frequency response of time domain anti-aliasing filter.]{
\psfrag{ot}{$\omega_t$}
\psfrag{xom}{$H_{\mathsf{aa}}(\omega_t)$}
\psfrag{w}{$v \rho$}
\psfrag{mw}{$-v \rho$}
\includegraphics[width=1.9in]{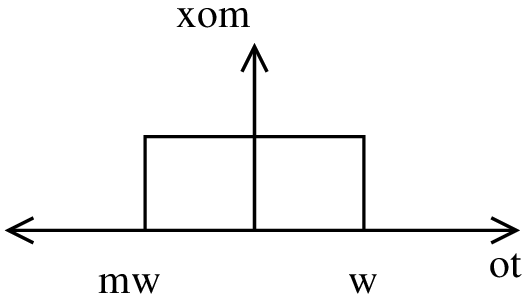}
\label{fig:aafil}
}
\subfigure[Frequency response of the induced sampling kernel is supported on $\{\omega:|\omega_x|\leq \rho\}$.]{
\psfrag{oy}{$\omega_y$}
\psfrag{ox}{$\omega_x$}
\psfrag{Hom0}{$H_{\sam}(\omega)=0$}
\psfrag{Hom1}{$H_{\sam}(\omega)=1$}
\psfrag{r}{$\rho$}
\psfrag{mr}{$-\rho$}
\includegraphics[width=1.9in]{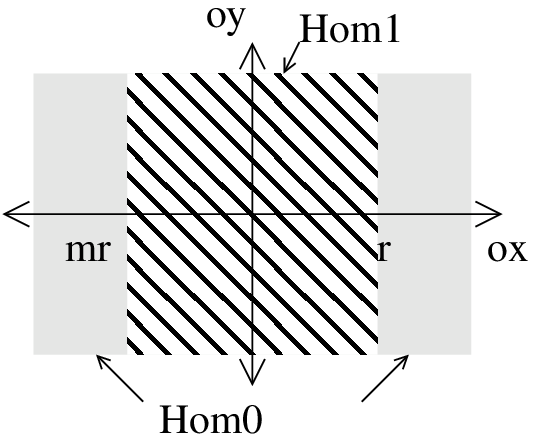}
\label{fig:hom2d}
}
\caption[Optional caption for list of figures]{Sampling a two-dimensional field using mobile sensors: Sensor trajectories, frequency response of time-domain filter, and the induced sampling kernel.}
\label{fig:subfigureExample}
\end{figure*}

\subsubsection{Two-dimensional field}\label{sec:samrecon2d}
In the case of a two-dimensional field $f(.)$, we consider sensors moving along equispaced parallel lines through space. For a given $\Omega$ the optimal orientation of these parallel lines can be computed as described in \cite{unnvet11TITb}. However, to simplify analysis we assume that the sensors are moving at a constant velocity $v$ along lines parallel to the $x$-axis spaced $\Delta_y$ units apart as illustrated in Figure \ref{fig:unifset}. The position of the $j$-th sensor at time $t$ is given by $(v t, j \Delta_y)$. Hence the $j$-th sensor is exposed to the signal $\tilde s_j(t)= \tilde f(v t, j \Delta_y)$. In the absence of noise, we know from (\ref{eqn:s0BW}) that these signals are bandlimited. Hence in the noisy scenario the signals $\tilde s_j(t)$ can be filtered and sampled uniformly, just like in the case of the one-dimensional field as we outlined in Section \ref{sec:samrecon1d}. In particular, if $\Omega = [-\rho, \rho] \times [-\rho, \rho]$, then it follows from (\ref{eqn:s0BW})  that the signal $s_0(t) = f(v t, j \Delta_y)$ is bandlimited to an interval of the form $[-\rho v,\rho v]$. Hence, like in Section \ref{sec:samrecon1d}, the filter $h_{\mathsf{aa}}(t) = v h(vt)$ can be used prior to uniform sampling. The frequency response of this ideal time-domain filter is shown in Figure \ref{fig:aafil}. Thus, at the output of the sampler we obtain uniform samples of
\[
s_j(t) = f(v t, j \Delta_y) + (w_j * h)(v t)
\]
where $w_j(x) = w(x, j \Delta_y)$ and $h$ is the filter defined in (\ref{eqn:hfilter}). Suppose that the samples are taken at time-intervals of $T < \frac{\pi}{v \rho}$ units. Let $\Delta_x = v T$. We further assume that the samples taken by the sensors are aligned with each other, and hence the collection of samples from all the sensors lie on a two-dimensional lattice of the form
$\{
(i \Delta_x, j \Delta_y): i,j \in \bZ
\}$
and can be expressed as
\begin{equation}
s_j(i T) = f(i \Delta_x, j \Delta_y) + (w_j * h)(i \Delta_x), \quad i,j \in \bZ. \label{eqn:mblsamples}
\end{equation}
Thus, results from classical sampling theory \cite{petmid62} can be used to estimate the field from these samples as:
\begin{eqnarray}
\hat f(x,y) &=& \sum_{i,j \in \bZ} \frac{\Delta_x \Delta_y \rho^2 s_j(i T)}{\pi^2}\sinc\left(\frac{\rho (x - i\Delta_x)}{\pi}\right)
\nonumber \\
&& \qquad \qquad \qquad \qquad \sinc\left(\frac{\rho (y - j\Delta_y)}{\pi}\right) \label{eqn:sincinterpnew}
\end{eqnarray}
provided $\Delta_x, \Delta_y < \frac{\pi}{\rho}$. The reconstruction given in (\ref{eqn:sincinterpnew}) is well-defined when the noise satisfies (\ref{eqn:noisemodel}) and is exact when noise is absent.

Using the notation $\mu[n] = s_j(iT)$ where $n = (i, j)^T$, the filtering and sampling relation of (\ref{eqn:mblsamples}) can be written as a two-dimensional convolution as follows:
\[
\mu[n] = (\tilde f \star h_{\sam})(\Lambda n), n \in \bZ^2,
\]
where $\star$ denotes two-dimensional convolution,
\[
\Lambda = \left(\begin{array}{cc}\Delta_x&0\\0&\Delta_y\end{array}\right),
\]
and $h_{\sam}$ represents the effective two-dimensional sampling kernel induced by the sampling trajectories of Figure \ref{fig:unifset} and the filtering operation of Figure \ref{fig:aafil} given by
\begin{equation}
h_{\sam}(x,y) = \frac{\rho}{\pi}\sinc\left(\frac{\rho x}{\pi}\right)\delta(y)\label{eqn:hsameff}
\end{equation}
where $\delta(.)$ represents the Dirac delta function. This kernel has the following representation in the Fourier domain as illustrated in Figure \ref{fig:hom2d}
\begin{eqnarray}
H_{\sam}(\omega ) = \left\{ \begin{array}{cc}1 & \mbox{for } 0 \leq |\omega_x| \leq {\rho}\\0& \mbox{else.}\end{array}\right. \label{eqn:sampkernmbl}
\end{eqnarray}
\begin{figure}
\centering
\psfrag{nu}{$\nu(r)$}
\psfrag{hs}{$h_{\sam}(.)$}
\psfrag{nr}{$\Lambda n: n\in\bZ^2$}
\psfrag{mn}{$\mu[n]$}
\psfrag{mr}{$\mu_s(r)$}
\psfrag{DC}{D/C}
\psfrag{hr}{$h_{\rec}(.)$}
\psfrag{nh}{$\hat \nu(r)$}
\psfrag{L}{$\Lambda$}
\includegraphics[width=3.4in]
{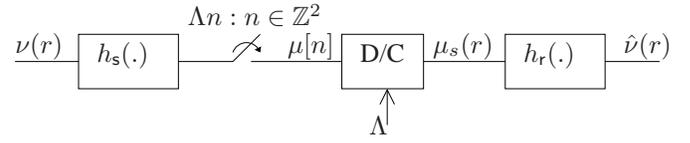}
\caption{Sampling and reconstruction setup.} \label{fig:system}
\end{figure}
Figure \ref{fig:system} shows the system of sampling and reconstruction where $\nu$ plays the role of $\tilde f$ and $\hat \nu$ the role of $\hat f$. In the figure the samples $\mu[n]$ passes through a discrete to continuous  converter to produce $\mu_s(r)$, an impulse stream in the continuous space given by
\[
\mu_s(r) = \sum_{n \in \bZ^2}\mu[n] \delta^2(r - \Lambda n)
\]
where $\delta^2$ represents the Dirac delta function in two dimensions. The reconstruction kernel $h_{\rec}(.)$ is the two-dimensional sinc-kernel
\[
h_{\rec}(x,y) = \frac{\Delta_x \Delta_y \rho^2}{\pi^2}\sinc\left(\frac{\rho x}{\pi}\right) \sinc\left(\frac{\rho y}{\pi}\right).
\]
The kernel has the following representation in the Fourier domain
\begin{eqnarray}
H_{\rec}(\omega ) = \left\{ \begin{array}{cc}\Delta_x\Delta_y & \mbox{for } 0 \leq |\omega_x|,|\omega_y| \leq {\rho}\\0& \mbox{else}\end{array}\right.\label{eqn:reconfil}
\end{eqnarray}
where $\omega = (\omega_x,\omega_y)$. In this representation the reconstructed field of (\ref{eqn:sincinterpnew}) is given by
\[
\hat f(r) = \sum_{n\in \bZ^2} \mu[n] h_{\rec}(r -\Lambda n), r \in \Re^2.
\]
We will use this new representation for simplifying the discussion in the rest of the paper. 

\subsection{Comparison with static sampling: Aliasing suppression for non-bandlimited fields}\label{sec:compwithstatnew}
Before proceeding to discuss bandlimited fields in detail, we now take a slight detour to consider the problem of sampling a non-bandlimited field in a noise-free setting. We know from classical sampling theory that sampling a signal at a rate less than the Nyquist rate leads to aliasing in the reconstructed signal, which is highly undesirable. In practice, while sampling a non-bandlimited signal in the time-domain, one typically employs an anti-aliasing filter to suppress the out-of-band portion of the signal. The ideal choice for the anti-aliasing filter and the reconstruction filter are ideal low-pass filters with cutoff frequencies given by half of the sampling rate. However, for sampling in the spatial domain, it is not possible to implement a spatial anti-aliasing filter in a static sensing setup. Nevertheless, in a mobile sampling scheme, an ideal time-domain filter can be used to reduce the amount of aliasing. In the one-dimensional case, as we argued in Section \ref{sec:samrecon1d}, this sampling procedure effectively implements an ideal anti-aliasing filter thereby suppressing aliasing completely. Thus the mobile sampling scheme is completely devoid of aliasing effects even while sampling a non-bandlimited field. Furthermore, the squared error in the reconstruction can be further reduced by simultaneously increasing the sampling rate and the bandwidth of the anti-aliasing filter.

\begin{figure}
\centering
\subfigure[Non-bandlimited spectrum.]{
\psfrag{Om}{$\Omega$}
\psfrag{Omh}{$\hat \Omega$}
\psfrag{SOI}{Spectrum of interest}
\psfrag{OOBE}{Out-of-band energy}
\includegraphics[width=1.4in]{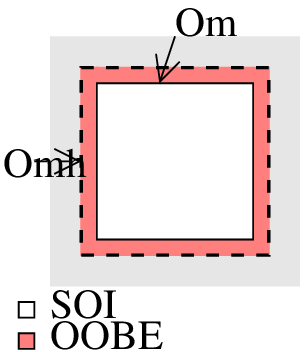}
\label{fig:nonBL}
}
\subfigure[Filtered spectrum.]{
\psfrag{Om}{$\Omega$}
\psfrag{Omh}{$\hat \Omega$}
\psfrag{SOI}{Spectrum of interest}
\psfrag{OOBE}{Out-of-band energy}
\includegraphics[width=1.4in]{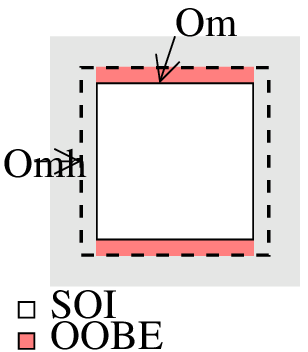}
\label{fig:nonBLfil}
}
\caption[Optional caption for list of figures]{Spectrum of a non-bandlimited two-dimensional field and its filtered version.}
\label{fig:nonBLandfil}
\end{figure}

\begin{figure*}
\centering
\subfigure[Sampled spectrum from static sampling.]{
\psfrag{Om}{$\Omega$}
\psfrag{Omh}{$\hat \Omega$}
\psfrag{NOS}{Non-overlapped spectrum}
\psfrag{B}{\Large \}}
\psfrag{OP}{Overlapped portions}
\includegraphics[width=1.5in]{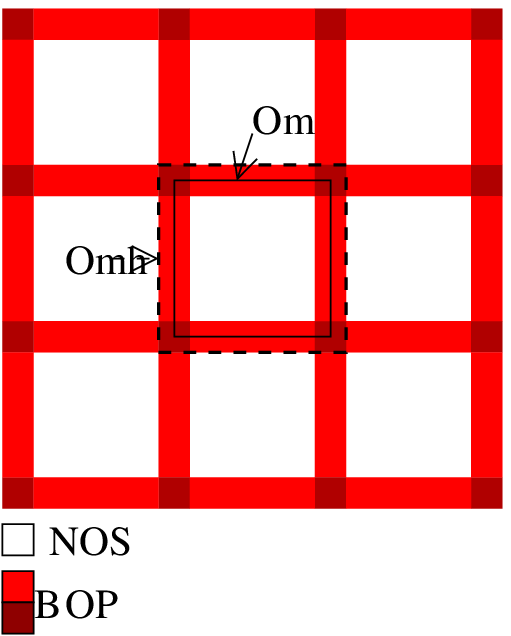}
\label{fig:statalias}
}
\subfigure[Reconstructed spectrum from static sampling.]{
\psfrag{Om}{$\Omega$}
\psfrag{UP}{Unaliased portion}
\psfrag{B}{\LARGE \}}
\psfrag{AP}{Aliased portions}
\includegraphics[width=1.4in]{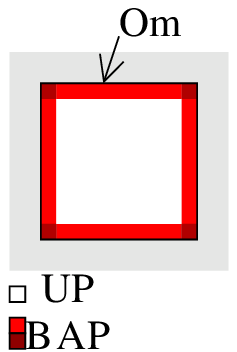}
\label{fig:stataliasrecon}
}
\subfigure[Sampled spectrum from mobile sampling with filtering.]{
\psfrag{Om}{$\Omega$}
\psfrag{Omh}{$\hat \Omega$}
\psfrag{NOS}{Non-overlapped spectrum}
\psfrag{OP}{Overlapped portions}
\includegraphics[width=1.5in]{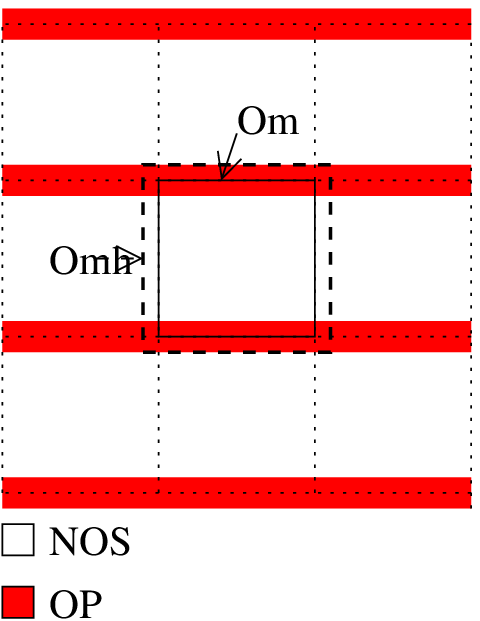}
\label{fig:mobalias}
}
\subfigure[Reconstructed spectrum from mobile sampling with filtering.]{
\psfrag{Om}{$\Omega$}
\psfrag{UP}{Unaliased portion}
\psfrag{AP}{Aliased portions}
\includegraphics[width=1.4in]{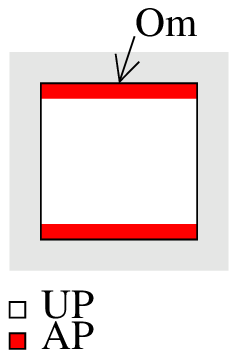}
\label{fig:mobaliasrecon}
}
\caption[Optional caption for list of figures]{Comparison of aliasing in static and mobile sampling schemes while sampling a two-dimensional field. The reconstructed field is aliased in two directions for the static scheme while only in one direction for the mobile scheme.}
\label{fig:aliasall}
\end{figure*}

The scenario is different in the two-dimensional case. Suppose we design the sampling scheme for fields bandlimited to $\Omega =[-\rho,\rho] \times [-\rho,\rho]$ and suppose that the observed noise-free field $\tilde f = f$ is in fact bandlimited to the bigger set\footnote{The assumption that the field is bandlimited to $\hat \Omega$ is only used to simplify the illustration. The advantages of filtering persist even when $f$ is not bandlimited.} $\hat \Omega = [-\hat \rho,\hat \rho]\times[-\hat \rho,\hat \rho]$ where $\hat \rho > \rho$. Such a field is shown in Figure \ref{fig:nonBL}. In a static sensing scheme this field is sampled without filtering and hence the sampled spectrum and reconstructed spectrum are aliased in both the $\omega_x$ and $\omega_y$ directions as shown in Figures \ref{fig:statalias} and \ref{fig:stataliasrecon}. In the mobile sampling case, however, we can use the anti-aliasing filter described in Section \ref{sec:samrecon2d}. In this case we know from the form of the effective sampling kernel in Figure \ref{fig:hom2d} that the out-of-band energy in the $\omega_x$ direction is filtered out as shown in Figure \ref{fig:nonBLfil}. Hence the sampled spectrum and reconstructed spectrum are aliased only in the $\omega_y$ direction as shown in Figures \ref{fig:mobalias} and \ref{fig:mobaliasrecon} respectively. Although the obtained reconstruction $\hat f$ is not completely devoid of aliasing, the squared error $\|f - \hat f\|_2^2$ in the reconstruction is significantly lower than the squared error in the reconstruction obtained via static sampling which is aliased both in the $\omega_x$ and $\omega_y$ directions.
%
%

Two observations are in order. Firstly, from the discussion above and Figure \ref{fig:aliasall} it is clear that the scheme of mobile sampling with filtering is effective in suppressing aliasing in the $\omega_x$ direction even when the field is not bandlimited in the $\omega_x$ direction. We discussed such an example in the introduction. The Fourier transform was supported on a semi-infinite set and the sampled spectrum is completely devoid of aliasing as illustrated in Figure \ref{fig:specBLwyall}. Such a complete suppression of aliasing is not possible with static sensing unless the field is bandlimited to a compact set in $\Re^2$. Secondly, it is possible to improve upon the mobile sampling scheme described here by employing mobile sampling along more complex sampling trajectories. For instance, in the example of Figure \ref{fig:nonBLandfil} if we replaced the sensors moving parallel to the $x$-axis with a similar set of sensors moving at constant velocities parallel to the $y$-axis then it would be possible to suppress aliasing in the $\omega_y$ direction by using ideal anti-aliasing filters as before. Now if we had both kinds of sensors, those moving along the $x$ direction and those moving along the $y$ direction, then the prefiltered samples from the former set of sensors are devoid of aliasing in the $\omega_x$ direction and the prefiltered samples from the latter set of sensors are devoid of aliasing in the $\omega_y$ direction. The samples from both these sets of sensors can be combined to obtain better aliasing suppression than that obtained in the Figure \ref{fig:mobaliasrecon}. A simple way to do this would be to first obtain two different reconstructions of the field, the first using the set of samples from the sensors moving along the $x$ direction and the other from the sensors moving along the $y$ direction. Now the aliased portions of each of these reconstructions can be filtered out and the resultant fields can be averaged to obtain a reconstruction with less aliasing than either original reconstruction.

\subsection{Comparison with static sampling: Noise suppression for bandlimited fields in noise}\label{sec:compwithstat}
We now study the reduction of out-of-band noise that can be obtained in the mobile sensing scheme using an anti-aliasing filter while sampling a bandlimited field in WSS noise satisfying (\ref{eqn:noisemodel}). The analysis is straightforward for one-dimensional fields. We provide a more detailed study for the two-dimensional case.

\subsubsection{One-dimensional field}
In a scheme of sampling a one-dimensional field using static sensors, one has access only to field measurements taken on a discrete set of points. Hence it is not possible to completely filter out the out-of-band noise in a static setup. However, if we use a sensor moving at constant velocity $v$, it is possible to filter out all the out-of-band noise as we showed in Section \ref{sec:samrecon1d}. We can quantify the advantage by comparing the noise variance in the reconstruction of (\ref{eqn:recon1d}) with the noise variance in the reconstruction of a static sampling scheme. Assume that field $f$ is bandlimited to $\Omega = [-\rho,\rho]$ and is sampled at the Nyquist rate. In the case of static sampling there is no anti-aliasing filter and hence the noise variance in the resulting reconstruction is given by
\begin{equation}
\sigma_{stat}^2 = \frac{1}{2\pi} \int_{-\rho}^\rho \sum_{n \in \bZ}\mathsf{S}_w(\omega -2 n \rho) d\omega \label{eqn:variancestat1d}
\end{equation}
where $\mathsf{S}_w(.)$ denotes the power spectral density (p.s.d.) of the noise process $w(.)$. In the mobile sampling case, however, the field is effectively filtered by an anti-aliasing filter given by (\ref{eqn:hfilter}) as in Section \ref{sec:samrecon1d} prior to sampling. Since the anti-aliasing filter prevents aliasing due to out-of-band noise, the total noise in the reconstruction of (\ref{eqn:recon1d}) is given by
\begin{equation}
\sigma_{m}^2 = \frac{1}{2\pi} \int_{-\rho}^\rho \mathsf{S}_w(\omega) d\omega. \label{eqn:variancembl1d}
\end{equation}
Clearly, we see that the contribution from the repetitions of $\mathsf{S}_w$ that appear in the static reconstruction of (\ref{eqn:variancestat1d}) is absent in the variance of noise in the reconstruction of (\ref{eqn:variancembl1d}) corresponding to the mobile sampling and filtering scheme of Section \ref{sec:samrecon1d}. Moreover, the contribution from the terms with $n \neq 0$ in (\ref{eqn:variancestat1d}) leads to aliasing in the reconstruction which is particularly undesirable.

\subsubsection{Two-dimensional field}\label{sec:2dfldcomp}
Consider a two-dimensional spatial field $f$ bandlimited to $\Omega = [-\rho,\rho]\times[-\rho,\rho] \subset \Re^2$. The static and mobile schemes for sampling the field can be represented as shown in Figure \ref{fig:system} where $\nu$ represents the input field being sampled, $h_{\sam}$ denotes the specific choice of the sampling kernel and $\Lambda$ is a $2 \times 2$ matrix that generates the spatial sampling lattice. In general, $\nu$ is a noisy version $\tilde f$ of the field $f$.

We consider a static sampling scheme that uses a rectangular sampling grid at the Nyquist sampling rate i.e., $\Delta_x = \Delta_y = \frac{\pi}{\rho}$, and we consider a mobile sampling scheme as in Section \ref{sec:samrecon2d} with the spacing $\Delta_y$ between the sensor trajectories equal to the Nyquist interval $\frac{\pi}{\rho}$. We consider two extreme choices for the low-pass filter employed prior to sampling in the mobile sensing scheme - the ideal low-pass filter (LPF) with a sinc response, and a more practically feasible filter, the box-filter whose impulse response is a rectangular pulse.

We first characterize the response of the sampling schemes to noise. Suppose the input $\nu$ in Figure \ref{fig:system} represents weak-sense stationary noise with power spectral density $\mathsf{S}_\nu( \omega), \omega \in \Re^2$. Then the p.s.d. of $\mu$ is given by,
\begin{equation*}
\mathsf{S}_\mu(e^{j\omega}) = \sum_{n \in \bZ^2}\frac{ \mathsf{S}_\nu(\Lambda^{-1}({\omega - 2\pi n}))|H_{\sam}(\Lambda^{-1}({\omega - 2\pi n}))|^2}{\Delta_x\Delta_y}. 
\end{equation*}
The output signal $\hat \nu$ is cyclostationary since its autocorrelation function is invariant to shifts by  $2\pi\Lambda^{-1}$. Its effective p.s.d. can be computed just as one computes the power spectra of linearly modulated signals (see, e.g., \cite[Sec. 3.4-2]{prosal08}) and is given by
\begin{equation}
\mathsf{S}_{\hat \nu}( \omega) = \frac{1}{\Delta_x\Delta_y}|H_{\rec}(\omega)|^2\mathsf{S}_\mu(e^{j \Lambda \omega}).\label{eqn:psdrecons}
\end{equation}
Below, we separately compute the noise variances in the case of static and the two cases of mobile sampling.

\medskip

\noindent \textbf{Static sampling}

\medskip

\noindent In the case of static sensing we do not have any spatial filtering and hence the sampling kernel is an impulse function
$
h_{\sam}(x,y) = \delta(x)\delta(y)
$
and is equivalently given in the Fourier domain by,
\begin{equation}
H_{\sam}(\omega) = 1 \mbox{ for all } \omega \in \Re^2. \label{eqn:sampkernstat}
\end{equation}
Since we assume Nyquist rate sampling we also have $\Delta_x = \Delta_y = \frac{\pi}{\rho}$. Substituting (\ref{eqn:sampkernstat}) in (\ref{eqn:psdrecons}) and integrating we obtain the following expression for the noise variance $\sigma_{stat}^2$:
\begin{equation}
\sigma_{stat}^2 = \frac{1}{4\pi^2} \int_\Omega  \sum_{n \in \bZ^2} \mathsf{S}_\nu(\omega -2 n \rho) d\omega. \label{eqn:variancestat}
\end{equation}
Since we are sampling at the Nyquist rate we do not observe any signal distortion in this case.

\medskip

\noindent \textbf{Mobile sampling with ideal low-pass filter}

\medskip

\noindent If the ideal sinc filter is employed as the LPF prior to sampling at the mobile sensors, the effective two-dimensional sampling kernel $h_{\sam}$ is given by (\ref{eqn:hsameff})
which is equivalently given in the Fourier domain by,
\begin{eqnarray}
H_{\sam}(\omega ) = \left\{ \begin{array}{cc}1 & \mbox{for } 0 \leq |\omega_x| \leq {\rho}\\0& \mbox{else.}\end{array}\right. \label{eqn:sampkernmbl}
\end{eqnarray}
We note that the filter $H_{\sam}$ is frequency limited in the x-direction. Further, since the spacing between the trajectories is equal to the Nyquist interval $\Delta_y = \frac{\pi}{\rho}$ it follows that as long as the sampling interval along the trajectories satisfies $\Delta_x < \frac{\pi}{\rho}$, we have the following expression for the noise variance $\sigma_{m1}^2$:
\begin{equation}
\sigma_{m1}^2 = \frac{1}{4\pi^2}\int_{\Omega}  \sum_{n \in \bZ} \mathsf{S}_\nu(\omega- (0, 2 n \rho)^T) d\omega. \label{eqn:variancembl}
\end{equation}
Comparing the expressions (\ref{eqn:variancestat}) and (\ref{eqn:variancembl}) we see that in the case of mobile sampling, only spectral shifts at lattice points along the x-axis contribute to the noise variance whereas in the case of static sampling spectral shifts from all points in the two-dimensional lattice contribute to the noise signal spectrum. The exact value of the reduction in noise variance obtained with mobile sampling can be computed if the true noise spectrum $\mathsf{S}_\nu(.)$ is known.


\medskip

\noindent \textbf{Mobile sampling with box filter}

\medskip

\noindent We now consider the case of mobile sampling with a box filter that is easier to implement. In this case we expect some distortion in the reconstruction because the filter response is not flat in the pass-band. The effective two-dimensional sampling kernel in this case is
$
h_{\sam}(x,y) = \rect(x)\delta(y)
$
where
\begin{eqnarray*}
\rect(x) := \left\{ \begin{array}{cc}
\kappa & |x| < \Delta_b\\
0 & \mbox{else}
\end{array} \right.
\end{eqnarray*}
where $2\Delta_b$ denotes the spatial width of the box-filter response and 
$
\kappa = (2\rho)^\half \left[ \frac{\Delta_b^2}{ \pi^2}\int_{-\rho}^\rho\sinc^2\left(\frac{\Delta_b \omega}{\pi}\right) d\omega\right]^{-\half}.
$
The sampling kernel is equivalently given in the Fourier domain by
\begin{eqnarray}
H_{\sam}(\omega ) =  \frac{\kappa \Delta_b}{\pi}\sinc \frac{\Delta_b \omega_x}{\pi}. \label{eqn:sampkernmbl2}
\end{eqnarray}
The choice of $\kappa$ ensures that
$
\int_\Omega H_{\sam}(\omega)^2 d\omega = 4\rho^2
$
which is consistent with the responses in (\ref{eqn:sampkernstat}) and (\ref{eqn:sampkernmbl}).

Since the box filter is a non-ideal LPF, we expect some distortion in the reconstructed field. As before, let us assume that $\Delta_x < \frac{\pi}{\rho}$ and that $\Delta_y = \frac{\pi}{\rho}$. Let $F$ denote the Fourier transform of the bandlimited field being sampled and $\hat F$ denote the Fourier transform of the reconstruction obtained by using an ideal reconstruction filter of the form (\ref{eqn:reconfil}). We have
\begin{eqnarray}
\hat F(\omega) =  \frac{\kappa \Delta_b}{\pi}F(\omega) \ \sinc(\frac{\Delta_b \omega_x}{\pi}), \, \omega \in \Omega. \label{eqn:mbl2reconorig}
\end{eqnarray}
Clearly, we see that the reconstructed field is a distorted version of the original field even in the absence of noise. The amount of distortion introduced for a given $F$ can be computed using relation (\ref{eqn:mbl2reconorig}). We can also quantify the variance $\sigma_{m2}^2$ of the noise in the reconstruction using relations (\ref{eqn:sampkernmbl2}) and (\ref{eqn:psdrecons}):
\begin{eqnarray}
\sigma_{m2}^2 &=& \frac{\kappa^2\Delta_b^2}{4\pi^4}\int_{\Omega}  \sum_{n \in \bZ^2} \mathsf{S}_\nu(\omega - (\frac{2 n_x \pi }{\Delta_x}, {2 n_y \rho})^T)\nonumber\\
&&\qquad \qquad \qquad \sinc^2(\frac{\Delta_b}{\pi}(\omega_x - \frac{2\pi n_x}{\Delta_x}))  d\omega. \nonumber \label{eqn:variancembl2orig}
\end{eqnarray}
Now if it is inexpensive to increase the sampling rate used by each moving sensor, we can let $\Delta_x \to 0$, whence we obtain,
\begin{eqnarray}
\sigma_{m2}^2 &=& \frac{\kappa^2\Delta_b^2}{4\pi^4}\int_{\Omega}  \sum_{n \in \bZ} \mathsf{S}_\nu(\omega - (0, {2 n \rho})^T)\sinc^2(\frac{\Delta_b \omega_x}{\pi})  d\omega. \nonumber\\\label{eqn:variancembl2}
\end{eqnarray}
If we now also allow $\Delta_b \to 0$, it is easy to see that $\frac{\kappa \Delta_b}{\pi} \to 1$. Hence the reconstruction in (\ref{eqn:mbl2reconorig}) becomes accurate and the expression in (\ref{eqn:variancembl2}) reduces to that in (\ref{eqn:variancembl}). This means that we have $\lim_{\Delta_b \to 0} \hat F(\omega) = F(\omega)$ for $\omega \in \Omega$, and $\lim_{\Delta_b \to 0} \lim_{\Delta_x \to 0} \sigma_{m2}^2 = \sigma_{m1}^2$.
This means that if the sensors oversample at high rates along their path and use a rectangular LPF with a short impulse response, we can recreate the performance with an ideal LPF and Nyquist sampling.

We note that if the noise spectral density  were white (i.e. $\mathsf{S}_\nu(\omega) = 1$ for all $\omega \in \Re^2$),  then the expressions for the variance in (\ref{eqn:variancestat}), (\ref{eqn:variancembl}) and (\ref{eqn:variancembl2orig}) tend to infinity. This is because unfiltered white noise samples have infinite variance. However, in practice environmental noise is never completely white. If the noise spectrum is flat with a bandwidth along each dimension equal to $a$ times the field bandwidth, then we obtain a reduction in noise variance by a factor of $a$ when we employ mobile sampling in place of static sampling, as shown below.

\begin{proposition}\label{prop:varratio}
Suppose the noise spectral density $\mathsf{S}_\nu(.)$ takes a value of unity on the set $[-a \rho, a \rho] \times [-a \rho, a \rho]$ and $0$ elsewhere for some $\rho > 0$. Then the variances in (\ref{eqn:variancestat}) and (\ref{eqn:variancembl}) decay as $O(a^2)$ and $O(a)$ respectively. In particular when $a$ is an odd number the variances reduce to
\[
\sigma_{stat}^2 = \frac{\rho^2 a^2}{\pi^2}, \qquad \mbox{ and } \qquad \sigma_{m1}^2 = \frac{\rho^2 a}{\pi^2}.
\]
\end{proposition}
Thus if the noise spectrum is flat with a bandwidth along each dimension equal to $a$ times the field bandwidth, then we obtain a reduction in noise variance by a factor of $a$ when we employ mobile sampling in place of static sampling.

\subsection{Noise suppression via oversampling on sensor paths}\label{sec:oversamp}
A unique aspect of a mobile sensing scheme is the fact that it is possible to sample at high rates along the paths of the mobile sensors. This is a significant advantage over static sensing, because in static sensing the sampling rate can be increased only by increasing the number of sensors which is, in general, more expensive than increase the density of samples taken by a moving sensor. Such oversampling helps in suppressing measurement noise, i.e., noise added to the discrete measurements after sampling. In the case of sampling a one-dimensional field bandlimited to $\Omega = [-\rho, \rho]$, as we discussed earlier, the Nyquist sampling rate is $\frac{v \rho}{\pi}$ where $v$ is the velocity of the sensor. It is known \cite[p. 136]{mal99} that, in the presence of additive zero-mean white measurement noise, oversampling by a factor of $k$ relative to the Nyquist rate leads to a reduction of noise variance in the reconstruction by a factor of $1/k$. Similarly, with non-linear processing quantization noise can be reduced by a factor of $1/k^2$ (see, e.g., \cite{thavet94}, \cite{cvevet98}, and references therein).

A similar noise reduction by oversampling can also be obtained while sampling a two-dimensional field. As in Section \ref{sec:2dfldcomp} let $\Omega = [-\rho,\rho]\times[-\rho,\rho]$ and assume that the sensors move along equispaced straight lines parallel to the x-axis and spaced $\Delta_y = \frac{\pi}{\rho}$ apart. The temporal Nyquist sampling rate along the lines is $\frac{v \rho}{\pi}$ where $v$ denotes the speed of the sensors. Now if the sensors take samples at $k$ times the Nyquist rate, the effective sampling operation can be interpreted as a measurement of the inner products of the field with vectors from $k$ disjoint orthogonal bases, as discussed in \cite[p. 136]{mal99}. Since these vectors form a tight frame with redundancy $k$, it follows via \cite[Prop. 5.3]{mal99} that oversampling by a factor of $k$ leads to a reduction in noise variance in the reconstruction by a factor of $1/k$ when sampling in the presence of additive zero-mean white measurement noise. It is possible that the known result \cite{thavet96} on the reduction of quantization noise by oversampling can also be extended to the two-dimensional case, but we do not consider such an extension here.

\subsection{Non-uniform sensor speeds}\label{sec:nonunifspeeds}
In practice it may not always be possible to ensure that the sensors move at a constant velocity. We now analyze the mobile sensing of a one-dimensional field $f$ bandlimited to $\Omega =[-\rho,\rho]$ using a sensor with time-varying speed. The analysis can be generalized to sensors moving along straight line trajectories in higher-dimensional fields, since the restriction of the higher-dimensional field to such trajectories is also bandlimited (see, e.g., \cite{bra99}). Suppose that the position $x(t)$ of the sensor at time $t$ is a known monotonically increasing function of time $t$. We have so far studied the case when $x(t)$ is an affine function, i.e., when the sensor has constant speed. In that case the signal $s_0(t)=f(x(t))$ is exactly bandlimited and hence can be sampled uniformly. For non-affine functions $x(t)$, the signal $s_0(t)=f(x(t))$ is not bandlimited. However, such a signal is also essentially bandlimited to some $\rho_0 < \infty$. The value of $\rho_0$ can be calculated as we show in the appendix. Hence we can obtain an approximate reconstruction of $s_0(t)$ by uniformly sampling an appropriately filtered version of it. Although this filtering operation is linear and time-invariant, there is still some distortion introduced because of the non-linearity in $x(t)$ which we now analyze. Let $h_{lp}(.)$ denote the impulse response of an ideal low-pass filter (LPF) employed prior to sampling. Let $\Delta_{lp}$ denote its $3$-dB spread in the temporal domain. Denote by $\tilde z(t)$ the output of the low-pass filter. We have
\begin{eqnarray}
\tilde z(t)= \int_\tau \tilde s(\tau) h_{lp}(t-\tau)d\tau=  z(t)+ \breve w(t)\label{eqn:lpfnonuniform}
\end{eqnarray}
with
\begin{eqnarray}
z(t)&=&  \int_\tau f(x(\tau)) h_{lp}(t-\tau)d\tau \nonumber\\
&=&  \int_x \frac{f(x) h_{lp}(t-T(x))}{v(T(x))}dx \label{eqn:varyingkernel}
\end{eqnarray}
where $v(t) = \frac{dx(t)}{dt}$ is the velocity function, $T(.) := x^{-1}(.)$ is the inverse function of $x(.)$, and $\breve w(t) = \int_\tau w(x(\tau)) h_{lp}(t-\tau)d\tau$. Thus at the output of the sampler, we get uniform samples $\{\tilde z_n:=\tilde z(t_n)\}$ at times $t_n := nT$. Two observations are in order. The spatial separation between two successive samples is approximately,
$
x_{n+1}-x_n \approx v(t_n) (t_{n+1} - t_n)
$
which means that samples are taken farther apart in space when the sensor is moving fast. We also note from (\ref{eqn:varyingkernel}) that the effective spatial spread of the sampling kernel while obtaining sample $\tilde z_n$ is given by $v(t_n) \Delta_{lp}$ which means that samples obtained while the sensor is moving fast are obtained via a broader effective sampling kernel in space. It is also clear from (\ref{eqn:varyingkernel}) that the sampling kernel is also scaled down by a factor proportional to the velocity. Now, since $\tilde z(t)$ is bandlimited, it can be reconstructed exactly from its uniformly spaced samples by sinc interpolation. Hence we can reconstruct an estimate for the field as $\hat f(x) := \tilde z(T(x))$. Clearly $\hat f(x)$ can be expressed as
$
\hat f(x) = z(T(x)) + \breve w(T(x))
$
where the first term $z(T(x))$ represents the contribution of the true field in the estimate and the second term represents the contribution of noise. We note that even in the absence of noise the reconstructed field is a distorted version of the field due to the non-linearity in $x(t)$.
The distortion in $z(T(x))$ can be quantified as follows:
\begin{eqnarray}
\int_\Re(f(x) - z(T(x)))^2 dx
&=& \int_\Re(s_0(t) - z(t))^2 v(t) dt  \nonumber\\
&\leq& \overline v\|s_0- z\|_2^2 
\end{eqnarray}
where $\overline v = \sup_t|v(t)|$  denotes the maximum speed of the sensor and $\|s_0- z\|_2^2$ denotes the total energy in $s_0(t)$ outside of the passband of the LPF. This suggests that the amount of distortion in the reconstruction can be reduced by increasing the bandwidth of the low-pass filter. This however comes at a cost of increasing the contribution of noise $w(.)$ in the reconstructed field $\hat f(.)$. As a heuristic one can use the effective bandwidth of the signal $s_0(t) = f(x(t))$ as the bandwidth of the low-pass filter. 


\section{Time-varying fields} \label{sec:tvarying}
We now consider the more general problem of sampling time-varying spatial fields using mobile sensors. We focus on time-varying fields in one-dimensional space - i.e., fields of the form $f(x,t)$ where $x \in \Re$ is a one-dimensional spatial parameter and $t$ denotes time. Our approach can be extended to time-varying fields in higher-dimensional spaces.


\subsection{Sampling and reconstruction}
Let $f(x,t)$ where $x,t\in\Re$, denote a time-varying field in one-dimensional space. Suppose that $f$ is bandlimited to $\Omega \subset \Re^2$. If $x(t)$ denotes the position of a moving sensor at time $t$, the signal seen by the sensor is given by $f(x(t),t)$. For a sensor moving at a constant speed $v$ the position is an affine function of time of the form $x(t) := u + v t$. Then it is clear from (\ref{eqn:s0BW}) that the signal $s_0(t) = f(x(t),t)$ is bandlimited to
\begin{equation}
\{v \omega_x + \omega_t: (\omega_x,\omega_t) \in \Omega\}.\label{eqn:s0BWmbl}
\end{equation}
Thus in the absence of noise, the signal $s_0(t)$ is bandlimited and can be exactly reconstructed by taking its samples at uniform intervals, as in the time-invariant case we considered in Section \ref{sec:samrecon2d}. In the presence of noise, an anti-aliasing filter with the appropriate bandwidth can be employed prior to sampling like in the time-invariant case. Furthermore, if the sensors are moving at non-uniform speeds then the signals are not exactly bandlimited but they can be approximated by bandlimited signals by following an approach similar to that in Section \ref{sec:nonunifspeeds}.

%

\begin{figure}
\centering
\subfigure[Sensor trajectories.]{
\centering
\psfrag{D}{$\Delta$}
\psfrag{Dc}{$\scriptstyle \Delta \cos \theta$}
\psfrag{th}{$\theta$}
\psfrag{t}{$t$}
\psfrag{x}{$x$}
\includegraphics[width=1.5in]
{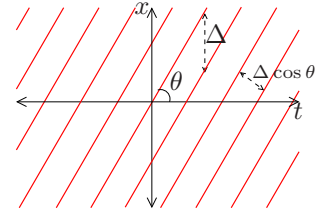}
\label{fig:timvartrajs}
}
\subfigure[Repetitions in sampled spectrum.]{
\centering
\psfrag{D}{$\Delta$}
\psfrag{Dc}{$\frac{2\pi }{\Delta \cos \theta}$}
\psfrag{nth}{$\scriptstyle \frac{\pi}{2} - \theta$}
\psfrag{O}{$\Omega$}
\psfrag{ox}{$\omega_x$}
\psfrag{ot}{$\omega_t$}
\psfrag{pt}{$\scriptstyle\rho_t$}
\psfrag{px}{$\scriptstyle\rho_x$}
\psfrag{mpt}{$\scriptstyle-\rho_t$}
\psfrag{mpx}{$\scriptstyle-\rho_x$}

\includegraphics[width=2in]
{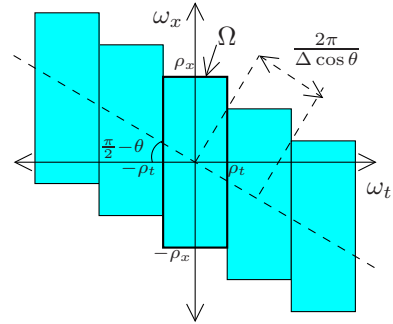}\label{fig:recspecvary}
}
\caption[Uniform configuration of moving sensors and resultant spectrum.]{Uniform configuration of moving sensors and resultant spectrum.}
\end{figure}

We consider a scheme of sampling using a uniform collection of mobile sensors moving with equal velocities and separated by a constant separation in space. Such a configuration of moving sensors is illustrated in Figure \ref{fig:timvartrajs}. Each line in the figure represents the position of an individual sensor as a function of time. The moving sensors are separated by a distance of $\Delta$ apart and move in the positive $x$ direction at a constant speed of $v$ represented by the slope $\tan \theta$ of the lines in the figure. From (\ref{eqn:s0BWmbl}) we know that the signal seen by each sensor is bandlimited to $\rho_t + v\rho_x$. We assume that the sensors sample in time at the temporal Nyquist rate of $\frac{\rho_t + v\rho_x}{\pi}$. We further assume that the samples taken by the various sensors are all synchronized in time such that the collection of all samples lie on a two-dimensional lattice. In this case, we know from classical sampling theory \cite{petmid62} that we can perfectly reconstruct any spatio-temporal field bandlimited to $\Omega$ from its values at these sample locations provided that the repetitions of its spectra do not overlap in the spectrum of the samples. Furthermore, since the temporal sampling rate is above the Nyquist rate, it can be shown that only repetitions along one direction need be considered. These repetitions are illustrated in Figure \ref{fig:recspecvary} for a rectangular set of the form $\Omega = [-\rho_x, \rho_x] \times [-\rho_t, \rho_t]$. A detailed explanation of this no-alias condition can be found in \cite{unnvet11TITb} for time-invariant fields.


We now explicitly compute the no-alias conditions for two specific choices of $\Omega$.

\subsubsection{Spatio-temporal field bandlimited to rectangular region}\label{sec:TVrect}
Suppose $\Omega = [-\rho_x, \rho_x] \times [-\rho_t, \rho_t]$ is a rectangular region. Under the sampling configuration described above, the condition to ensure that there is no aliasing in the field reconstruction is that the repetitions shown in Figure \ref{fig:recspecvary} do not overlap. This means that $\Delta$ should satisfy either
\[
\frac{2\pi }{\Delta} > 2\rho_x \, \qquad \mbox{ or } \qquad \frac{2\pi }{\Delta}\tan \theta > 2\rho_t.
\]
Since $\tan \theta$ represents the velocity of the sensors, the above condition is equivalent to the following requirement on the spatial separation between adjacent moving sensors:
\begin{equation}
\Delta < \pi \max \left\{\frac{v}{\rho_t}, \frac{1}{\rho_x} \right\}. \label{eqn:1dtvlimit}
\end{equation}
Thus $\pi \max \left\{\frac{v}{\rho_t}, \frac{1}{\rho_x} \right\}$ is the maximum admissible spatial separation between adjacent sensors.

\subsubsection{Wave field}\label{sec:TVwave}
\begin{figure}
\centering
\subfigure[Wave field spectrum]{
\centering
\psfrag{O}{$\Omega$}
\psfrag{rt}{$\rho_t$}
\psfrag{rx}{$\rho_x$}
\psfrag{ox}{$\omega_x$}
\psfrag{ot}{$\omega_t$}
\includegraphics[width=1.4in]
{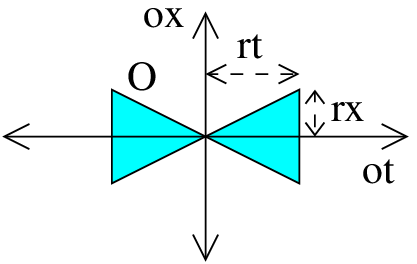}
\label{fig:wavfldspec}
}
\subfigure[Sampled wave field spectrum]{
\centering
\psfrag{D}{$\Delta$}
\psfrag{Dc}{$\frac{2\pi }{\Delta \cos \theta}$}
\psfrag{nth}{$\scriptstyle \frac{\pi}{2} - \theta$}
\psfrag{th}{$\theta$}
\psfrag{O}{$\Omega$}
\psfrag{ox}{$\omega_x$}
\psfrag{ot}{$\omega_t$}
\includegraphics[width=1.75in]
{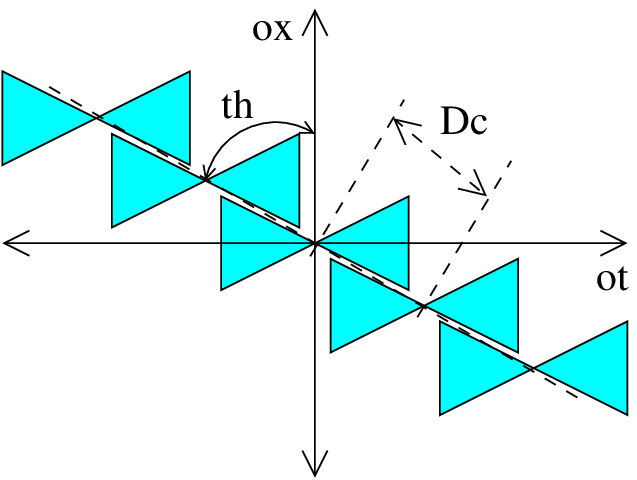}
\label{fig:samwavfldspec}
}
\caption[]{Far-field spectrum of a bandlimited source and its sampled version from samples taken by the mobile sensors of Figure \ref{fig:timvartrajs}.}
\end{figure}

We now consider a time-varying field with a non-rectangular frequency spectrum. Suppose we are interested in reconstructing the spatio-temporal wave field along a line. Assume that the field is produced by bandlimited sources located far from the region of interest. In this setting, we can use the far-field approximation to study the spectrum of the bandlimited field. It was shown in \cite{ajdsbavet06} that the spectrum of such a field is approximately supported on the region shown in Figure \ref{fig:wavfldspec}. Here $\rho_t$ is the bandwidth of the source signals and $\rho_x = \frac{\rho_t}{c}$ where $c$ denotes the speed of propagation of the wave. Now suppose that we sample the field using moving sensors with trajectories shown in Figure \ref{fig:timvartrajs} under the sampling configuration described before. Then, as before the condition required to ensure that there is no aliasing is that the spectral repetitions in Figure \ref{fig:samwavfldspec} do not overlap. It follows from the figure that for sensor velocities $v < c$ a sufficient condition on the spacing $\Delta$ to ensure that the spectral repetitions do not overlap is that
$\displaystyle
-\frac{2\pi}{\Delta}< -2 \rho_x + \frac{\rho_x}{\rho_t} \frac{2\pi}{\Delta}\tan \theta,
$
or equivalently
\begin{equation}
\Delta < \frac{\pi}{\rho_x}(1 + \frac{v}{c}).\label{eqn:1dtvlimitwavefield}
\end{equation}
A similar analysis can also be performed for sampling a wave field over two-dimensional space using moving arrays of sensors. The sampling of such a field using an array of sensors is described in \cite{col04}. These ideas can be extended to the case of mobile sampling by following an approach like the one we described in this section.

%

\subsection{Comparison with static sampling}
In Section \ref{sec:compwithstat} we noted the advantages of mobile sampling over static sampling obtained by using an anti-aliasing filter to limit the contribution of out-of-band noise while sampling and reconstructing time-invariant fields. For sampling time-varying fields, however, such an advantage is not as significant. Consider filtering and sampling a one-dimensional time-varying field using sensors moving according to the configuration depicted in Figure \ref{fig:timvartrajs}. In this case, we are essentially filtering along the lines in the $t$-$x$ plane shown in Figure \ref{fig:timvartrajs}. We note that it is possible to filter over time even with static sensors. This would amount to filtering along lines parallel to the $t$-axis in Figure \ref{fig:timvartrajs}. Thus the only difference between filtering in the mobile and static sensing cases is in the direction of filtering in the $t$-$x$ plane. Hence the relative advantages of the two schemes would depend on the spectral characteristics of the additive noise. However, mobile sensing offers a different sort of advantage over static sensing: In some situations, we can get a reduction in the spatial density of sensor deployment required while using mobile sensors, as we show below.

Consider a time-varying field bandlimited to a rectangular region $[-\rho_x, \rho_x] \times [-\rho_t, \rho_t]$ as in Section \ref{sec:TVrect}. We know from classical sampling \cite{petmid62} that for sampling with static sensors the maximum spacing allowed between adjacent sensors is $\frac{\pi}{\rho_x}$. Comparing with the spacing requirement in the mobile setting given in (\ref{eqn:1dtvlimit}), it follows that when the mobile sensors are moving at a speed $v > \frac{\rho_t}{\rho_x}$, the inter-sensor spacing can be increased by a factor of
$
\frac{v \rho_x}{\rho_t}.
$
In other words, this means that for a given length of the spatial region of interest, we can reduce the number of sensors required by a factor
$
\frac{\rho_t}{v\rho_x}.
$
The advantage is more significant when $v \gg \frac{\rho_t}{\rho_x}$, i.e., for spatio-temporal fields that vary slowly in time and at fast rate over space. This matches with the intuition that slowly varying fields are easier to track using mobile sensors. However, as the speed $v$ is increased, the required temporal rate of sampling given by the Nyquist rate, $\frac{\rho_t + v \rho_x}{\pi}$, also increases. Hence, in short, by using moving sensors we can reduce the spatial density of sensors at the cost of increasing their temporal sampling rates.


Now consider the scenario of sampling a wave field along a line located in the far field of bandlimited sources as in Section \ref{sec:TVwave}. For sampling with static sensors the maximum spacing allowed between adjacent sensors is again $\frac{\pi}{\rho_x}$. Hence it follows from (\ref{eqn:1dtvlimitwavefield}) that the inter-sensor spacing can be increased by a factor of $(1+\frac{v}{c})$ when we employ sensors moving at speed $v$. This observation suggests that for wave field reconstruction we get a significant improvement in the sensor spacing using mobile sensing only when the sensors can move at a speed of the order of the speed of wave propagation in the medium.




\begin{figure}
\centering
\includegraphics[width=3in]{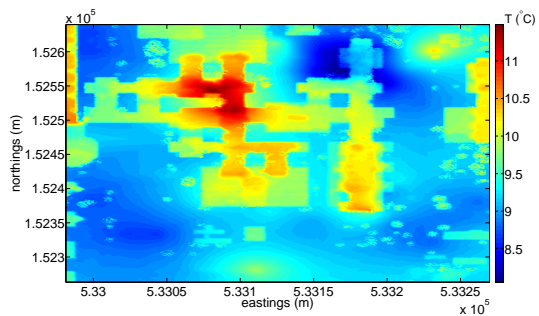}
\caption{Spatial temperature field on EPFL campus.}
\label{fig:temfield}
\end{figure}

\begin{figure}
\centering
\includegraphics[width=3in]{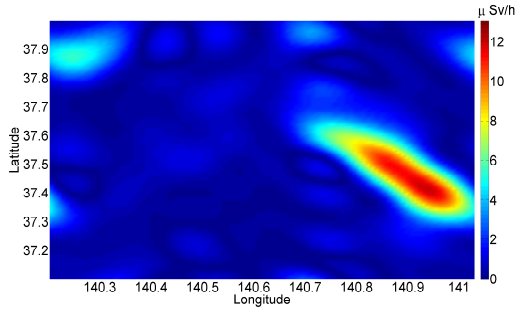}
\caption{Bandlimited approximation of the radiation field near Fukushima.}
\label{fig:radfield}
\end{figure}

\begin{table}
    \caption{Percentage root-mean-square errors with various schemes.}
    \centering
    \begin{tabular}{ | p{2.2cm} | p{0.9cm} | p{0.9cm} | p{2.8cm} |}
    \hline
    Data type & Static sensing & Mobile sensing & Mobile sensing with oversampling\\ \hline
    Temperature & 0.53\% & 0.45\%  & 0.42\% (\mbox{{\small  no filter}})\\ \hline
    Bandlimited radiation (SNR $20$ dB)& 9.9\% & 1.5\% & 1.5\% (\mbox{{\small with filter}})  \\ \hline
    \end{tabular}
    \label{tab:com}
\end{table}
\section{Simulations: Merits of mobile sensing}\label{sec:sims}
We simulated the static and mobile sampling schemes for measuring the surface temperature field on a portion of the EPFL campus. For the true temperature field we used the readings obtained from \cite{nadetal09} as illustrated in Figure \ref{fig:temfield}. For static sensing, we considered sensors on a rectangular grid. For mobile sensing we assumed that the sensors move parallel to the $x$-axis and apply an anti-aliasing filter prior to sampling, like in Section \ref{sec:samrecon2d}. They take samples at the same points on the rectangular grid as in the static case. As seen in Figure \ref{fig:temfield} the field has sharp variations in space and hence is not bandlimited. Thus we expect some aliasing in the reconstruction obtained via sinc interpolation from samples of such a field. In Table \ref{tab:com} we list the percentage root-mean-square errors in the reconstructed fields, defined as $\frac{\|\hat f - f\|_2}{\|f\|_2}\times 100$. The last column represents the performance obtained with mobile sensing assuming that the sensors measure the field at all points on their paths without any filtering. As the values in the table indicate, mobile sampling outperforms static sampling, and oversampling along the trajectories improves the performance further. The filtering operation in the mobile sampling scheme reduces the amount of aliasing in the samples leading to a reduction in the reconstruction error. We note that the temperature field is not truly bandlimited and hence the performance gains are more modest than what could be expected if the field were truly bandlimited.

We also simulated the same schemes for sampling and reconstructing a truly bandlimited field in noise. For the true field shown in Figure \ref{fig:radfield} we used a bandlimited approximation to the spatial radiation field around the site of the Fukushima nuclear accident on 11 March 2011. The radiation levels in this field were measured in units of micro-Sieverts per hour ($\mu \mbox{Sv}/h$) at various positions during the months of July - September, 2011, and are available online at \cite{safecast}. We considered the sampling of a noisy version of this field with the noise spectrum as described in the statement of Proposition \ref{prop:varratio}, with the ratio of the sides $a = 40$ and estimated the percentage root-mean-square errors $\displaystyle \frac{\sqrt \mathsf{E}[\|\hat f - f\|_2^2]}{\|f\|_2}\times 100$ in the reconstruction. The values of the errors shown in Table \ref{tab:com} suggest that the reduction in the error obtained with mobile sensing is more significant than that was seen for the temperature field. We also see that the ratio of the errors under the static and mobile reconstruction schemes is approximately $a^\half$ as expected by the result of Proposition \ref{prop:varratio}. In this example we allow filtering in the oversampling scheme since the field of interest is bandlimited. From the last column of Table \ref{tab:com} we see that there is no improvement in accuracy with oversampling. This is expected since there is no advantage in increasing the sampling rate beyond the Nyquist rate.

\section{Conclusion and future work}\label{sec:conc}
In this paper we have studied strategies for sampling and reconstructing a bandlimited spatial field using moving sensors, including both time-invariant and time-varying fields. We highlighted and quantified the advantages of mobile sensing over classical static sensing, both in theory and through simulations. Our results for time-invariant fields clearly demonstrate the following advantages when using a time domain anti-aliasing filter together with a mobile sensor:
\begin{romannum}
\item For non-bandlimited fields: Anti-aliasing filtering with mobile sampling suppresses aliasing in the direction of motion. For one-dimensional fields, higher sampling rates yield lower distortion in the reconstruction.
\item For bandlimited fields in noise: Anti-aliasing filtering with mobile sampling suppresses noise in the direction of motion. This prevents aliasing in the direction of motion. Sampling at the Nyquist rate is sufficient.
\end{romannum}
In the latter case we quantified the SNR improvement of the mobile sampling scheme over the static sampling scheme. For time-varying fields, we demonstrated the improvement in sampling density that can be obtained by using mobile sensors.

In our analysis of mobile sensing of time-invariant fields in $\Re^2$ we considered only trajectories composed of a set of equispaced parallel lines. The analysis of SNR in the reconstruction can be generalized to higher dimensional spaces and to more general configurations of straight line trajectories like the ones studied in \cite{unnvet11TITb}. In practice one may be forced to use non-linear sensor trajectories. In such cases, a possible approach would be to approximate the paths by straight line segments and then use the appropriate anti-aliasing filters to reduce spatial anti-aliasing as we did for the non-uniform speed sensing example of Section \ref{sec:nonunifspeeds}. However, quantifying the noise suppression in such cases would be more complex. For general trajectories, it would be interesting to study the tradeoff between the SNR improvement and the \emph{path density metric} of the trajectories introduced in \cite{unnvet11TITb}. Similar extensions are also relevant for studying time-varying fields.


For making mobile sampling schemes practical, one would also need to consider the effects of mobility on the field and on the sensing process. It is possible that the physical process of moving the sensor through the field may affect the characteristics of the field or introduce noise and irregularities in the sensing process. These effects must also be taken into account to completely characterize the advantages of mobile sensing over static sensing in practice.

\section*{Acknowledgements}
This research was supported by ERC Advanced Investigators Grant: Sparse Sampling: Theory, Algorithms and Applications – SPARSAM – no 247006.

\appendix[Bandwidth of time-warped signal $s_0(t) = f(x(t))$]
Consider a piecewise-affine function of the form
\begin{equation}
x_1(t) = \sum_{k = 1}^K (u_k + v_k t)\clI\{t \in [t_k, t_{k+1})\}.\label{eqn:piecewiselinear}
\end{equation}
where $t_k < t_{k+1}$, and $0 \leq v_k \leq \overline v$ where $\overline v$ denotes the maximum speed of the sensor. Let $\Delta_k := t_{k+1} - t_k$, $\Delta := \min_k \Delta_k$ and $\overline t_k := \frac{t_{k+1} + t_k}{2}$. We know that the Fourier transform of $s_1(t) := f(x_1(t))$ is given by
\begin{equation}
S_1(\xi) = \sum_{k = 1}^K \left[ \frac{1}{v_k}e^{\frac{ja_k\xi}{v_k}} F\left(\frac{\xi}{v_k}\right) *_\xi e^{{-j\xi\overline t_k}}\Delta_k\sinc \left(\frac{\xi \Delta_k}{2\pi}\right)\right]\label{eqn:S0spectrum}
\end{equation}
where $F(.)$ denotes the Fourier transform of the field $f(.)$, and the operation $ *_\xi$ denotes convolution with respect to $\xi$. From the structure of the Fourier transform we can argue as in \cite{domarvet12} that the effective bandwidth of $s_1$ is given by
\begin{equation}
\max_k [v_k \rho + \frac{1}{\Delta_k}] \leq \overline v \rho + \frac{1}{\Delta}.
\end{equation}
We now study how the Fourier transform of the observed signal gets modified by a slight deviation from the piecewise affine trajectory. Suppose the trajectory is given by
\begin{equation}
x(t) = \sum_{k = 1}^K (u_k + v_k t + \epsilon \tilde x_k(t))\clI\{t \in [t_k, t_{k+1})\}. \label{eqn:epsapproxtraj}
\end{equation}
Let $\rho_{\tilde x}$ denote the maximum low-pass bandwidth of all $\tilde x_k$. Assume that $f$ is twice differentiable. Then we have by Taylor's approximation that for $t \in [t_k, t_{k+1})$,
\begin{eqnarray*}
\lefteqn{s_0(t) = f(x(t))}\\ &=& f(u_k + v_k t + \epsilon \tilde x_k(t)), \quad \mbox{for } t \in [t_k, t_{k+1})\\
&=& f(u_k + v_k t)  + f'(u_k + v_k t) \epsilon \tilde x_k(t) + \BigO(\epsilon^2)
\end{eqnarray*}
Using $S_0(\xi)$ to denote the spectrum of $s_0(t)$ it follows that for small $\epsilon$ we have
\begin{eqnarray}
S_0(\xi) - S_1(\xi) &=&\epsilon \sum_{k = 1}^K \left\{ \frac{j \xi}{v_k^2} e^{\frac{ja_k\xi}{v_k}} F\left(\frac{\xi}{v_k}\right) *_\xi \tilde X_k(\xi) *_\xi\right.\nonumber \\
&& \left.e^{\frac{-j\xi\Delta_k}{2}}\Delta_k\sinc \left(\frac{\xi \Delta_k}{2\pi}\right)\right\} + \BigO(\epsilon^2).\label{eqn:S0approxerror}
\end{eqnarray}
Thus, we can argue that for small $\epsilon$ the difference between the Fourier transforms of $s_0(t)$ and $s_1(t)$ is given by a term proportional to $\epsilon$ over frequencies in the range
\[
|\xi| \leq \max_k [v_k \rho_f + \rho_{\tilde x_k} + \frac{1}{\Delta_k}] \leq \overline v \rho_f + \rho_{\tilde x} + \frac{1}{\Delta}
\]
and only by terms of order $\BigO(\epsilon^2)$ for other frequencies. By considering higher order terms in the Taylor series expansion, it follows that the difference between Fourier transforms of $s_0(t)$ and $s_1(t)$ is given by a term of order $\BigO(\epsilon^{m+1})$ over frequencies outside of the range
\[
|\xi| \leq \overline v \rho_f + m \rho_{\tilde x} + \frac{1}{\Delta}.
\]

\bibliographystyle{IEEEtran}
\bibliography{TomrefsSP}
\end{document}